\documentclass[a4paper,twocolumn,11pt,unpublished]{quantumarticle}
\pdfoutput=1
\usepackage[utf8]{inputenc}
\usepackage[english]{babel}
\usepackage[T1]{fontenc}
\usepackage{amsmath}
\usepackage{hyperref}
\hypersetup{colorlinks=true,linkcolor=blue,urlcolor=blue,citecolor=blue}
\usepackage[numbers,sort&compress]{natbib}
\usepackage{physics}
\usepackage{amssymb}
\usepackage{amsthm}    % for proof of theorem
\newtheorem{corollary}{Corollary} %for corollary
\usepackage{tikz}
\usepackage{lipsum}

\newtheorem{theorem}{Theorem}

\begin{document}

\title{Spatial entanglement between two quantum walkers with exchange symmetric coins}

\author{Ibrahim Yahaya Muhammad}
\affiliation{Theoretical and Computational Physics (TCP) Group, Department of Physics, Faculty of Science, King Mongkut's University of Technology Thonburi, Bangkok 10140, Thailand}
\affiliation{Theoretical and Computational Science Center (TaCS), Faculty of Science, King Mongkut's University of Technology Thonburi, Bangkok 10140, Thailand}
%\orcid{0000-0003-4176-2995}
\author{Tanapat Deesuwan}
\email{tanapat.dee@kmutt.ac.th}
\affiliation{Theoretical and Computational Physics (TCP) Group, Department of Physics, Faculty of Science, King Mongkut's University of Technology Thonburi, Bangkok 10140, Thailand}
\affiliation{Theoretical and Computational Science Center (TaCS), Faculty of Science, King Mongkut's University of Technology Thonburi, Bangkok 10140, Thailand}
\affiliation{Quantum Computing and Information Research Centre (QX), Faculty of Science, King Mongkut’s University of Technology Thonburi, Bangkok 10140, Thailand}
%\orcid{0000-0003-1162-5041}
\author{Sikarin Yoo-Kong}
\affiliation{The Institute for Fundamental Study (IF), Naresuan University, Phitsanulok 65000, Thailand}
%\email{sikariny@nu.ac.th}
\author{Suwat Tangwancharoen}
\affiliation{Theoretical and Computational Physics (TCP) Group, Department of Physics, Faculty of Science, King Mongkut's University of Technology Thonburi, Bangkok 10140, Thailand}
\affiliation{Theoretical and Computational Science Center (TaCS), Faculty of Science, King Mongkut's University of Technology Thonburi, Bangkok 10140, Thailand}
%\email{suwat.tan@kmutt.ac.th}
\author{Monsit Tanasittikosol}
\email{monsit.tan@kmutt.ac.th}
\affiliation{Theoretical and Computational Physics (TCP) Group, Department of Physics, Faculty of Science, King Mongkut's University of Technology Thonburi, Bangkok 10140, Thailand}
\maketitle

\begin{abstract}
We investigate how the initial and final exchange symmetries between the two-coin states influence the spatial entanglement dynamics between the two corresponding quantum walkers. Notably, when the initial state is anti-symmetric and the final measurement on the coins yields symmetric outcomes, all the initial entanglement will be transferred to the spatial degrees of freedom, regardless of when the coins are measured. Conversely, if the final outcomes are anti-symmetric, the spatial entanglement exhibits damped oscillation with a period ($T$) being inversely proportional to the coin operator parameter ($\theta$). These behaviours are reversed for symmetric initial states. Moreover, we also observe the same spatial entanglement damping regardless of the initial state when the post-selected results lack symmetry. Our findings reveal how symmetries affect the entanglement dynamics in quantum walks, offering potential insights for applications in quantum technology.
\end{abstract}

% In the \texttt{twocolumn} layout and without the \texttt{titlepage} option a paragraph without a previous section title may directly follow the abstract.
% In \texttt{onecolumn} format or with a dedicated \texttt{titlepage}, this should be avoided.

% Note that clicking the title performs a search for that title on \href{http://quantum-journal.org}{quantum-journal.org}.
% In this way readers can easily verify whether a work using the \texttt{quantumarticle} class was actually published in Quantum.
% If you would like to use \texttt{quantumarticle} for manuscripts not yet accepted in Quantum, or not even intended for submission to Quantum, please use the \texttt{unpublished} option to switch off all Quantum related branding and the hyperlink in the title.
% By default, this class also performs various checks to make sure the manuscript will compile well on the arXiv.
% If you do not intend to submit your manuscript to Quantum or the arXiv, you can switch off these checks with the \texttt{noarxiv} option.
% On the contrary, by giving the \texttt{accepted=YYYY-MM-DD} option, with \texttt{YYYY-MM-DD} the acceptance date, the note ``Accepted in Quantum YYYY-MM-DD, click title to verify'' can be added to the bottom of each page to clearly mark works that have been accepted in Quantum.

\section{\label{Intro}Introduction}
Since it was first studied by Aharonov \textit{ et al.} in $1993$ \cite{aharonov1993quantum}, quantum walk has become an interesting research area in quantum information and quantum computation. It is radically different from its classical counterpart due to quantum properties, such as, superposition and entanglement. The power of quantum walk lies in its wide range of potential applications. It is shown in \cite{childs2009universal,lovett2010universal,childs2013universal,Kendon2020HowTC,gong2021quantum} that quantum walk can implement universal quantum computation and some important quantum computing/communication tasks \cite{shenvi2003quantum,santha2008quantum,wang2017generalized,chen2019quantum,giordani2021entanglement,sahu2024a}. It has also been used to simulate biological systems \cite{acunto2021,varsamis2023}, quantum materials \cite{karafyllidis,dadras2018quantum} and even some phenomena in high-energy physics,
for example, neutrino oscillations \cite{sahu2024b}, and the dynamics of a spin 1/2 particle in a certain warped geometry \cite{angles2022}. Moreover, physical implementation of quantum walk in some physical systems such as trapped ions \cite{schmitz2009quantum,zahringer2010realization}, nuclear magnetic resonance \cite{ryan2005experimental}, cold atoms \cite{preiss2015strongly,mugel2016topological}, integrated photonics \cite{schreiber2010photons,peruzzo2010quantum}, and superconducting qubits \cite{flurin2017observing,zhou2019protocol}, shows its promising role in future quantum technologies. An introductory overview and a review of quantum walk can be found in \cite{kempe2003quantum,venegas2012quantum}.

Apart from its application side, the intrinsic behavior of quantum walk itself, particularly the entanglement among the walkers and the coins, also attracts significant research interest. Studies have explored various scenarios to understand and potentially enhance this entanglement. For instance, it was shown both numerically \cite{carneiro2005entanglement} and analytically \cite{abal2006quantum} that the dynamics of the entanglement between one walker and one coin oscillate around an asymptotic value. Additionally, the entanglement between two coins and one walker has been studied \cite{annabestani2010}. Moreover, an enhancement of the two coins and one walker entanglement has been achieved by incorporating artificial magnetic fields \cite{yalccinkaya2015} and dynamical disorder \cite{zeng2017} into the quantum walks. Recent work \cite{naves2023} employed entangled coins to enhance the two coins and one walker entanglement. Furthermore, spatial entanglement between two walkers sharing a single coin has been investigated in \cite{venegas2009quantum, alles2012maximal} through coin-basis measurements to generate entanglement between the walkers. Additionally, various studies \cite{omar2006quantum, berry2011two, carson2015entanglement, stefanak2006meeting, pathak2007quantum, chandrashekar2012entanglement, singh2019accelerated, goyal2010spatial} have explored entanglement between two systems by considering a walker and its coin as one system. However, spatial entanglement with two walkers sharing two coin states with post-selection on the coin states remains unexplored.

In this paper, starting with a two-coin state with either symmetric or anti-symmetric exchange symmetry, we study the dynamics of spatial entanglement between two walkers undergoing discrete-time quantum walk on a one-dimensional lattice, with post-selection on the coin states at the end of the process. Particularly, we investigate how the initial and final symmetries of the two coins affect the entanglement between the walkers when the total step of the walk is varied. The structure of this paper is as follows. In Section \ref{sec2}, we provide a concise overview of quantum operations for a one-walker, one-coin quantum walk and extend them to a scenario with two walkers and two coins. In Section \ref{sec3}, starting with the two-entangled-coin state initially, we analyze the entanglement dynamics between two walkers, where we present and discuss our main findings. The final section \ref{conc} provides the conclusion of our study.
\section{\label{sec2}Quantum walk}
Let us first start with an overview of the one-walker and two-walker quantum walk together with the mathematical notations which will be used throughout the text.
\subsection{\label{sec2.1}One walker with one coin}
To introduce the discrete-time quantum walk with one walker in one dimension, we imagine the motion of a walker on a one-dimensional lattice. The position of the walker lattice point $i$ is assigned by the state $\ket{i}$, where $i$ $\in$ $\mathbb{Z}$, which belongs to the Hilbert space of the walker $\mathcal{H}_w$. The walker needs to toss a coin and move according to the outcome. The Hilbert space of the coin $\mathcal{H}_c$ is defined by a basis consisting of two orthonormal states: $\{\ket{\uparrow}, \ket{\downarrow}\}$. The total state of the system (walker+coin) then belongs to the product space $\mathcal{H}=\mathcal{H}_c\otimes\mathcal{H}_w$.
There are 3 operations involved in the quantum walk as follows:
\begin{itemize}
\item \textbf{Operation 1}: Apply the coin operator $\hat{C}_{1}(\theta)$ to the coin state. The coin operator is defined as
\begin{align}
\hat{C}_{1}(\theta) = \mqty(\cos\theta& \sin\theta \\ \sin\theta & -\cos\theta),
\label{eq1}
\end{align}
where $\theta$ is the angle that determines how biased the coin is. For example, $\theta=\pi/4$ means we have an unbiased coin, while $\theta=n\pi,$ where $n$ is an integer means only one side of the coin always show up, i.e., we have a maximally biased coin.
\item	\textbf{Operation 2}: Apply the shift operation $\hat{S}_{1}$ to the walker state. The shift operator is defined as
\begin{align}
\hat{S}_{1} = \ket{\uparrow}\bra{\uparrow}\otimes \displaystyle\sum_{i}\ket{i+1}\bra{i} \nonumber\\
+ \ket{\downarrow}\bra{\downarrow}\otimes \displaystyle\sum_{i}\ket{i-1}\bra{i}.
\label{eq2}
\end{align}
Note that this operation entangles the walker and the coin. The first two operations can be combined into a single unitary operator:
\begin{eqnarray}
    \hat{U}_{1} = \hat{S}_1(\hat{C}_1\otimes \hat{I}).
\label{eq3}
\end{eqnarray}
where $\hat{I}$ is the identity operator acting on the position state of the walker. The operator $\hat{U}_1$ will be applied to the system as many times as one needs before proceeding to the last operation. We will say the $n^{th}$ steps of the walk is done after $\hat{U}_1$ is applied to the system $n$ times.
\item \textbf{Operation 3}: Measure the state of the coin. The measurement basis is usually formed by the following operators,
\begin{eqnarray}
    \hat{M}_{\uparrow} = \ket{\uparrow}\bra{\uparrow} \;,\; \hat{M}_{\downarrow} = \ket{\downarrow}\bra{\downarrow}.
\label{eq4}
\end{eqnarray}
Due to the prior entanglement generated via the shift operation, the state of the walker will collapse and become a classical spatial distribution that correlates with the outcome of the coin measurement.
\end{itemize}
%%%%%%%%%%%%%%%%%%%%%%%%%%%%%%%%%%%%%%%%%%%%%%%%%%%%%%%%%%%%%%%%%%%%%%%%%%%%%%%%%
\subsection{\label{sec2.2}Two walkers with two identical coins}
Let us now consider two walkers, each of which \emph{separately} but \emph{identically} evolves according to the previously discussed one walker protocol with respect to each own coin. The Hilbert space of the total system, in this case, is the product space $\mathcal{H}$ given by
$
    \mathcal{H}=\mathcal{H}_1\otimes\mathcal{H}_2=(\mathcal{H}_c\otimes\mathcal{H}_p)_1\otimes(\mathcal{H}_c\otimes\mathcal{H}_p)_2,
$
where $\mathcal{H}_1$ is the combined Hilbert space of walker 1 and coin 1, and $\mathcal{H}_2$ is the combined Hilbert space of walker 2 and coin 2. The coin operator in this case is
\begin{eqnarray}
    \hat{C}(\theta) = \hat{C}_1(\theta)\otimes\hat{C}_2(\theta). 
\label{eq5}
\end{eqnarray}
We only consider the case where the angles for the two coins operators are the same. The shift operator is
\begin{align}
\hat{S} &= \hat{S}_1\otimes\hat{S}_2\nonumber\\
&= \ket{\uparrow\uparrow}\bra{\uparrow\uparrow}\otimes \sum_{ij}\ket{i+1,j+1}\bra{i,j}\nonumber\\
&+ \ket{\uparrow\downarrow}\bra{\uparrow\downarrow}\otimes \sum_{ij}\ket{i+1,j-1}\bra{i,j}\nonumber \\
&+ \ket{\downarrow\uparrow}\bra{\downarrow\uparrow}\otimes\sum_{ij}\ket{i-1,j+1}\bra{i,j}\nonumber\\
&+ \ket{\downarrow\downarrow}\bra{\downarrow\downarrow}\otimes\sum_{ij}\ket{i-1,j-1}\bra{i,j}.
\label{eq6}
\end{align}
Thus, the combination of these two operators gives the evolution operator in the form,
\begin{eqnarray}
    \hat{U}(\theta) = \hat{S}(\hat{C}(\theta)\otimes I).
\label{eq7}
\end{eqnarray}
Furthermore, the measurement basis considered in this case are composed of the following operators,
\begin{align}
\hat{M}_{\uparrow\uparrow}&=\ket{\uparrow\uparrow}\bra{\uparrow\uparrow}\;,\;\hat{M}_{\uparrow\downarrow}=\ket{\uparrow\downarrow}\bra{\uparrow\downarrow}\nonumber\\
\hat{M}_{\downarrow\uparrow}&=\ket{\downarrow\uparrow}\bra{\downarrow\uparrow}\;,\;\hat{M}_{\downarrow\downarrow}=\ket{\downarrow\downarrow}\bra{\downarrow\downarrow} \;.
\label{eq8}
\end{align}
Note that all of these are basically two copies of the one walker quantum walk operators that apply to the two walkers separately. The fact that all the operations considered are ``local'' implies that the entanglement between the two walkers cannot be increased by these operations.
%%%%%%%%%%%%%%%%%%%%%%%%%%%%%%%%%%%%%%%%%%%%%%%%%%%%%%%%%%%%%%%%%%%%%%%%%%%%%%%%%
\section{\label{sec3}Our settings and results}
Our setting is a system of two walkers with two identical coins with the following initial state.% Suppose that the initial state of the system is a product between the state of the two coins and the state of the two walkers that can be written in the following form,
\begin{eqnarray}
\ket{\psi}_0 = \ket{Bell}_c\otimes\ket{0,0}_w,
\label{eq10}
\end{eqnarray}
where $\ket{Bell}_c$ is the state of the two coins which is in one of the Bell states ($\ket{\Psi^{\pm}}=\frac{\ket{\uparrow,\downarrow}\pm\ket{\downarrow,\uparrow}}{\sqrt{2}}$, $\ket{\Phi^{\pm}}=\frac{\ket{\uparrow,\uparrow}\pm\ket{\downarrow,\downarrow}}{\sqrt{2}}$). We can classify these initial coin states according to the exchange symmetry into two types, symmetric ($\ket{\Psi^{+}}$, $\ket{\Phi^{\pm}}$) and anti-symmetric state ($\ket{\Psi^{-}}$). For the initial positions of the two walkers, we choose them to be the same at the origin for convenience. In fact, without loss of generality, we can also choose them to be at any other point, even if each one of them is in a superposition state, as long as they are identical. With this particular setting, if $\ket{Bell}_c$ is symmetric, Eq.\,\ref{eq10} will represent the states of two bosons, while if it is anti-symmetric, it will describe the state of two fermions.

Each walker then undergoes an identical but separate discrete-time quantum walk, as discussed in the previous section. To investigate the entanglement dynamics between the two walkers, we will use the entanglement entropy as the entanglement measure
\begin{eqnarray}
    E(\ket{\psi}) = - \sum_j \lambda_j \log_2 \lambda_j, 
\label{eq9}
\end{eqnarray}
where $\lambda_j$ are the eigenvalues of the reduced density matrix for subsystem 1 or subsystem 2.
%%%%%%%%%%%%%%%%%%%%%%%%%%%%%%%%%%%%%%%%%%%%%%%%%%%%%%%%%%%%%%%%%%%%%%%%%%%%%%%%%
\subsection{Theorem}
Let us first proof the following two Theorems, which we will use to analyze our results.
\begin{theorem}
Given a bipartite system, a completely positive map that is symmetric under the exchange of the order of the input state cannot change the symmetric property of the whole system.
\label{t1}
\end{theorem}
\begin{proof}
Let us define an exchange operator $\hat{P}$, which is simultaneously Hermitian and unitary operator. Now, we apply $\hat{P}$ to the state $\ket{a,b}$. 
%\\\\
\begin{eqnarray*}
\hat{P}\ket{a,b}=\pm \ket{b,a}; \quad \hat{P}^2=\hat{I}
\end{eqnarray*}
%\\\\
The value $+1$ means that the state is symmetric, and $-1$ means the state is anti-symmetric. We also consider a completely positive operator $\hat{\varphi}_{ab}$, where $\hat{\varphi}_{ab}=\hat{\varphi}_{ba}$. In other words, the way this operator works does not depend on the order of the input states.
% \begin{equation*}
%     \hat{P}\ket{A,B}=e^{i\theta} \ket{B,A}; \quad \hat{P}\hat{P}^\dagger=\hat{I},
% \end{equation*}
Then
\begin{eqnarray}
    \hat{P}\hat{\varphi}_{ab}\ket{\psi}_{ab} &=& \hat{P}\hat{\varphi}_{ab}\hat{P}^2\ket{\psi}_{ab} \nonumber \\
%    where \nonumber \\
    \hat{P}\hat{\varphi}_{ab}\ket{\psi}_{ab} &=& \hat{\varphi}_{ba} \hat{P}\ket{\psi}_{ab} \nonumber \\
   (\hat{P}\hat{\varphi}_{ab} - \hat{\varphi}_{ba} \hat{P})\ket{\psi}_{ab} &=& 0 \nonumber \\
    \big[\hat{P},\hat{\varphi}_{ab}\big]\ket{\psi}_{ab} &=& 0.
\end{eqnarray}
\end{proof}
Therefore, $\big[\hat{P},\hat{\varphi}_{ab}\big]=0$, which indicates that the operator $\hat{\varphi}_{ab}$ does not change the symmetric property of the system.
\begin{corollary}
Two identical local operations cannot change the symmetric property of a system.  \label{corollary}
\end{corollary}
\begin{proof}
Let us define two local operator $\Omega_a$ and $\Omega_b$, where $\Omega_a=\Omega_b=\Omega$ and $\Omega_{ab}=\Omega_a\otimes \Omega_b=\Omega\otimes\Omega$. It is clear that $\Omega_{ab}=\Omega_{ba}$, and therefore $\Omega_{ab}$ does not affect the exchange symmetry of the system according to Theorem \ref{t1}.
\end{proof}
\begin{theorem}
If the initial state of a walker is pure, then the reduced density matrix of that walker after post-selection on any orthonormal basis of the corresponding coin state always has at most rank 2.
\label{t2}
\end{theorem}
\begin{proof}
Let us consider our scenario but only on Alice's side. Without loss of generality, we choose Alice's initial state to be
\begin{eqnarray}
    \rho_{A} = (q_1\ket{\uparrow}\bra{\uparrow}+q_2\ket{\downarrow}\bra{\downarrow})\otimes \ket{w}\bra{w}
\end{eqnarray}
where $q_1+q_2=1$ and $\ket{w}$ is the initial state of the walker. Note that this does not mean that the walker has to occupy only one definite point initially. A superposition of points is allowed as long as the state is pure. Now, if we run the quantum walk by applying the operator $\hat{U}_1$ in Eq.\,\ref{eq3}, the whole state will change to
\begin{eqnarray}
\hat{U}_1\rho_{A}\hat{U}_{1}^{\dagger} &=& \hat{U}_1(q_1\ket{\uparrow}\bra{\uparrow}\otimes\ket{w}\bra{w})\hat{U}_{1}^{\dagger} \nonumber\\
&&+ \hat{U}_1(q_2\ket{\downarrow}\bra{\downarrow}\otimes\ket{w}\bra{w})\hat{U}_{1}^{\dagger}\nonumber\\
&=& q_1(\lambda_1\ket{c}\ket{w_1}+\lambda_2\ket{c^\perp}\ket{w_2}) \nonumber\\
&&\otimes (\lambda_{1}^{*}\bra{c}\bra{w_1} +\lambda_{2}^{*}\bra{c^\perp}\bra{w_2}) \nonumber\\
&& + q_2(\lambda_3\ket{c}\ket{w_3} + \lambda_4\ket{c^\perp}\ket{w_4}) \nonumber\\
&&\otimes (\lambda_{3}^{*}\bra{c}\bra{w_3}+\lambda_{4}^{*}\bra{c^\perp}\bra{w_4}) \nonumber\\
\end{eqnarray}
where $\{\ket{c},\ket{c^\perp}\}$ is an arbitrary orthogonal basis of the Hilbert space of the coin and $\lambda_i$ are the corresponding probability amplitudes. Suppose we post-select the coin state $\ket{c}\bra{c}$, the total state becomes
\begin{eqnarray}
\rho_{A}^{c} &= \ket{c}\bra{c}\otimes \rho_{w}^{c},
\end{eqnarray}
where $\rho_w^{c}=\frac{q_1|\lambda_1|^2\ket{w_1}\bra{w_1}+q_2|\lambda_3|^2\ket{w_3}\bra{w_3}}{q_1|\lambda_1|^2+q_2|\lambda_3|^2}$ is the corresponding walker state.  Note that $\ket{w_1}\bra{w_1}$ and $\ket{w_3}\bra{w_3}$ are individually \textit{pure state} and a pure state always has rank 1. Using the relation
\begin{eqnarray}
rank(M+N) \leq rank(M) + rank(N),
\end{eqnarray}
we can conclude that
\begin{eqnarray}
rank(\rho_{w}^{c}) \leq 2
\end{eqnarray}
Note that this is also true if we post-select in $\ket{c^\perp}\bra{c^\perp}$ state. In other words, the walker effectively becomes a two-dimensional system or a qubit after post-selection.
\end{proof}
%%%%%%%%%%%%%%%%%%%%%%%%%%%%%%%%%%%%%%%%%%%%%%%%%%%%%%%%%%%%%%%%%%%%%%%%%%%%%%%%%
\subsection{Entanglement between the two walkers}
Now, we will examine how the symmetries between the two coins, both at the beginning of the walk and at the end, affect the dynamics of the spatial entanglement. Since our initial coin state is fixed to be one of the Bell states and there are clearly two types of exchange symmetry associated to them, we will separately analyse the two cases of symmetric and anti-symmetric initial coin state separately. We will also choose the set of ``local'' measurement operators described in Eq.\,\ref{eq8} as the measurement basis before considering a specific ``non-local'' one. The reason we are interested in post-selection is because, in many situations, post-selection can reveal some hidden correlations between the systems, which then may be used to perform some interesting tasks that would be impossible otherwise, for example, quantum teleportation or quantum key distribution. The reason we focus on a local measurement basis first is because it cannot increase entanglement between the systems, and, in many applications, only this kind of measurements are allowed.
%%%%%%%%%%%%%%%%%%%%%%%%%%%%%%%%%%%%%%%%%%%%%%%%%%%%%%%%%%%%%%%%%%%%%%%%%%%%%%%%%
\subsubsection{\label{sec3.2}Anti-symmetric initial Coin State (ACS)}
In this case, the initial state of the system is given by
\begin{eqnarray}
\ket{\psi}_{0}=\ket{\Psi^{-}}\otimes\ket{0,0}.
\label{eq15}
\end{eqnarray}
%This state is anti-symmetric. 
We will perform the quantum walk with the unitary operator $\hat{U}(\pi/4)$ in Eq.\,\ref{eq7}. These choices of the initial positions and the angle parameter $\theta$ are only for convenience. The results do not actually depend on them as long as the initial states of the two walkers are the same and $\hat{U}(\theta)$ satisfies Theorem \ref{t1}. Now we will consider, for example, the state after the second step of the quantum walk.
\begin{widetext}
\begin{equation}
\ket{\psi}_{2} = \frac{1}{2\sqrt{2}}\big[\ket{\uparrow\uparrow}(\ket{0,2}-\ket{2,0})+\ket{\uparrow\downarrow}(\ket{2,-2}+\ket{0,0})-\ket{\downarrow\uparrow}(\ket{0,0}+\ket{-2,2})+\ket{\downarrow\downarrow}(\ket{0,-2}-\ket{-2,0})\big].
\label{eq16}
\end{equation}
\end{widetext}
We can see that the state of the system after the second step is still anti-symmetric under the exchange of coins and walkers. This is because $\hat{U}$ acts symmetrically on both sides. In fact, with Theorem \ref{t1}, it is clear that the state $\ket{\psi}_{n}$ after any $n^{th}$ step must be anti-symmetric. Now, let us perform a local measurement on each coin. The state of the walkers for the post-selected outcomes of $\hat{M}_{\uparrow\uparrow}$, $\hat{M}_{\uparrow\downarrow}$, $\hat{M}_{\downarrow\uparrow}$ and $\hat{M}_{\downarrow\downarrow}$ are
\begin{eqnarray}
 \ket{\psi}_{2,w}^{\hat{M}_{\uparrow\uparrow}} &=& \frac{1}{\sqrt{2}}(\ket{0,2}-\ket{2,0}),\nonumber\\ 
 \ket{\psi}_{2,w}^{\hat{M}_{\uparrow\downarrow}} &=& \frac{1}{\sqrt{2}}(\ket{2,-2}+\ket{0,0}),\nonumber\\
 \ket{\psi}_{2,w}^{\hat{M}_{\downarrow\uparrow}} &=& -\frac{1}{\sqrt{2}}(\ket{0,0}+\ket{-2,2}),\nonumber\\ \ket{\psi}_{2,w}^{\hat{M}_{\downarrow\downarrow}} &=& \frac{1}{\sqrt{2}}(\ket{0,-2}-\ket{-2,0}).
 \label{eq17}
\end{eqnarray}
We can see that the states of the two walkers are anti-symmetric if we post-select one of the symmetric outcomes ($\ket{\uparrow\uparrow}$ or $\ket{\downarrow\downarrow}$). However, for the other two cases where the outcome does not have any symmetry ($\ket{\uparrow\downarrow} or \ket{\downarrow\uparrow}$), the states of the two walkers will also lack symmetry. Again, according to Theorem \ref{t1}, this must also be true even if the two-coin state is measured and post-selected after any $n^{th}$ step of the walk.

\begin{figure}[ht]
\centering
\includegraphics[width=1\columnwidth]{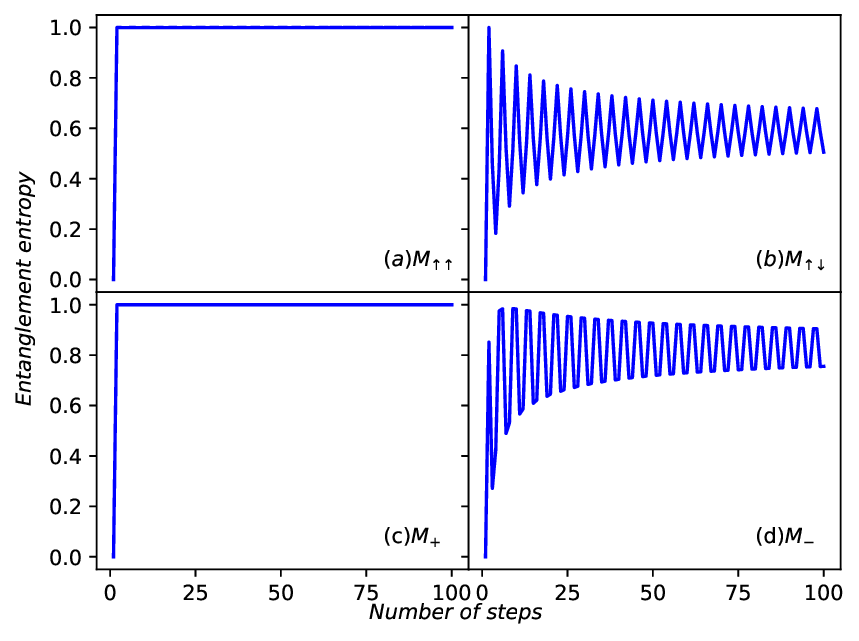}
\caption{\label{fig1} Entanglement between two walkers for \textit{ACS} case (as in Eq.~\ref{eq15}) using operator $\hat{U}(\pi/4)$ for different post-selection (a) $\hat{M}_{\uparrow\uparrow}$, (b) $\hat{M}_{\uparrow\downarrow}$, (c) $\hat{M}_{+}$ (d) $\hat{M}_{-}$.}
\end{figure}

Let us consider the anti-symmetric walker states first. The general form of a bipartite anti-symmetric state can be written as
\begin{eqnarray}
    \ket{\xi}=\sum_{ij}\alpha^{\prime}_{ij}(\ket{a_i,a_j}-\ket{a_j,a_i}),
    \label{24}
\end{eqnarray}
where $\{\ket{a_i}\}$ is a set of vectors that forms an orthonormal basis for the Hilbert space of walker $i$ and $\alpha^{\prime}_{i,j}$ is the corresponding probability amplitude. Using Theorem \ref{t2}, the reduced density matrix of a single walker in Eq.\,\ref{24} can only have rank 2 at most. In other words, the state is equivalent to a qubit. Therefore, only one pair would remain in Eq.\,\ref{24}. Without loss of generality, we choose that pair to be of the form $\frac{1}{\sqrt 2}(\ket{a,b}-\ket{b,a})$. This form is equivalent to $\ket{\Psi^-}$, which is the only anti-symmetric state whose reduced density matrix has at most rank 2. This explains the reason why we always have the same value of spatial entanglement when we post-select the $\ket{\uparrow\uparrow}$ or $\ket{\downarrow\downarrow}$ result, as shown numerically in Fig.\,\ref{fig1}(a). Note that, since the entanglement within this whole system stems from the initial Bell state of the coins, this result indicates that the entanglement is totally transferred from the coins to the walkers and this is the maximum achievable value within this setting.

Next, let us consider the other two cases where the reduced states of the walkers lack symmetry. Even though, Theorem \ref{t2} is still satisfied in these cases, the walkers' state cannot be a Bell state and would not have constant maximum entanglement as in the previous case. In fact, our numerical results, plotted in Fig.\,\ref{fig1}(b), show that the spatial entanglement will change with the number of steps in an underdamped-like manner and eventually reaching an asymptotic value of $0.5884$. %We also observe that the oscillation period $(T)$, the number of walking steps $(n)$, and the angle parameter of the coin operator $(\theta)$ are related via the equation: $T=n\pi/\theta$ (cf. Fig.~\ref{fig4}(b)).
This indicates that there are entanglement exchange between different degrees of freedom within the system, which will eventually settle down and reach an equilibrium in a long time limit. Note that, this behaviour is really similar to the entanglement between the coin and the walker discovered earlier in \cite{carneiro2005entanglement}. In fact, we also discover a numerical relationship between the period of entanglement oscillation and the angle parameter of the coin operator, but we will describe this result later in the last section of this paper before the conclusion.

The results that we have got so far reveal an interesting relationship between initial and final symmetries of the coins and the dynamics of the entanglement within the systems. It clearly shows that only if the final states of the coins are symmetric when the initial states are anti-symmetric, the final states of the walkers would be anti-symmetric and the initial entanglement would be completely transfer to the walker. There would be no transient fluctuation and equilibration-like behaviour, unlike the cases where the final states of the coins have no symmetry. %It should be emphasised that this interesting role of symmetry only becomes manifest in the setting like ours.
This leads to an interesting question: What would happen to the spatial entanglement if we make it such that all the four possible outcomes from the final coin measurements possess some symmetries? If only symmetry is really behind such the undamping behaviour, all the four possible post-selection results should now all be undamping. To verify this, let us change $\hat{M}_{\uparrow\downarrow}$ and $\hat{M}_{\downarrow\uparrow}$ in the measurement basis of Eq.\,\ref{eq8} to $\hat{M}_{+}=\frac{1}{2}(\ket{\uparrow\downarrow}+\ket{\downarrow\uparrow})(\bra{\uparrow\downarrow}+\bra{\downarrow\uparrow})$ and $\hat{M}_{-}=\frac{1}{2}(\ket{\uparrow\downarrow}-\ket{\downarrow\uparrow})(\bra{\uparrow\downarrow}-\bra{\downarrow\uparrow})$ which, together with $\hat{M}_{\uparrow\uparrow}$ and $\hat{M}_{\downarrow\downarrow}$ still forms a complete basis. Now our measurement basis consists of non-local measurement operators $\hat{M}_{+}$ and $\hat{M}_{-}$. %However, these non-local measurement operators may come with the concern of increasing entanglement of the system and also most of the time is very difficult task to perform in the real-world application. 
We may call this new basis the triplet-singlet basis. Note that, the only reason we change to this basis for now is to investigate further into the role of the coin symmetry in determining the dynamics of the spatial entanglement. In the context of quantum random walk, we are still restricted to local measurements only.

Using the new basis, we will consider the state after the second step of the quantum walk Eq.\,\ref{eq16}, the state of the walkers for the post-selection outcome of $\hat{M}_{+}$ and $\hat{M}_{-}$ are given by
\begin{eqnarray}
 \ket{\psi}_{2,w}^{\hat{M}_{+}} &=& \frac{1}{\sqrt{2}}(\ket{2,-2}-\ket{-2,2}),\nonumber\\
 \ket{\psi}_{2,w}^{\hat{M}_{-}} &=& \frac{1}{\sqrt{6}}(\ket{2,-2}+2\ket{0,0}+\ket{-2,2}).
 \label{31}
\end{eqnarray}
The walkers' state corresponding to $\hat{M}_{+}$ is anti-symmetric while $\hat{M}_{-}$ is symmetric as expected. Figure \ref{fig1}(c) shows that the entanglement dynamics is constant for $\hat{M}_{+}$ post-selection, while the entanglement dynamics of $\hat{M}_{-}$ post-selection is underdamping as shown in Fig.\,\ref{fig1}(d) with an asymptotic value of $0.8143$. This result shows that symmetry alone does not guarantee the undamping behaviour. From observation of the previous results, it is clear that, rather than only require that the final coin states possess some certain symmetries, the condition should actually be that the final states of the walkers are anti-symmetric. In other words, the symmetries of the initial and the final coin states must be opposite. We will proceed further to check the validity of this claim by changing the symmetry of the initial coin state.

%%%%%%%%%%%%%%%%%%%%%%%%%%%%%%%%%%%%%%%%%%%%%%%%%%%%%%%%%%%%%%%%%%%%%%%%%%%%%%%%%
\subsubsection{\label{sec3.3}Symmetric initial Coin State (SCS)}
\begin{figure}[ht]
\centering
\includegraphics[width=1\columnwidth]{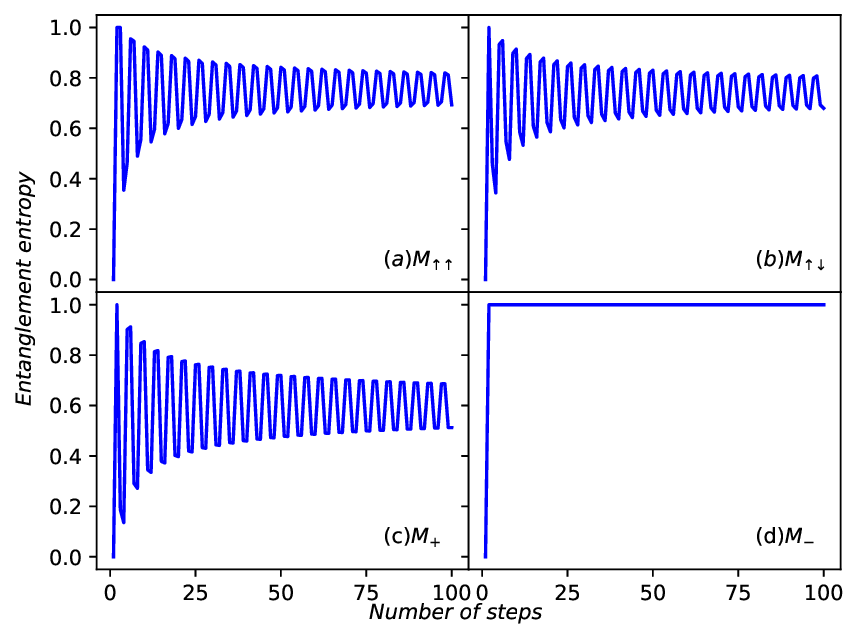}
\caption{\label{fig2} Entanglement between two walkers for \textit{SCS} case (as in Eq.~\ref{eq18}) using operator $\hat{U}(\pi/4)$ for different post-selection (a) $\hat{M}_{\uparrow\uparrow}$, (b) $\hat{M}_{\uparrow\downarrow}$, (c) $\hat{M}_{+}$ (d) $\hat{M}_{-}$.}
\end{figure}

Let us now change the initial state of the coins to be symmetric, given by
\begin{eqnarray}
\ket{\psi}_0=\ket{\Psi^{+}}\otimes\ket{0,0}.
\label{eq18}
\end{eqnarray}
The other two cases of symmetric initial state will also be discussed later on. We can use Theorem \ref{t1} to show that the state $\ket{\psi}_{n}$ after the $n^{th}$ steps is symmetric, for instance, the state after the second step of the quantum walk using the unitary operator $\hat{U}(\pi/4)$ is
\begin{widetext}
\begin{equation}
\ket{\psi}_{2} = \frac{1}{2\sqrt{2}}\big[\ket{\uparrow\uparrow}(\ket{2,2}-\ket{0,0})+\ket{\uparrow\downarrow}(\ket{2,0}+\ket{0,-2})+\ket{\downarrow\uparrow}(\ket{0,2}+\ket{-2,0})+\ket{\downarrow\downarrow}(\ket{0,0}-\ket{-2,-2})\big],
\label{eq19}
\end{equation}
\end{widetext}
which is symmetric under the exchange of both coins and walkers. Now, let us measure the state after the second step. The walkers' state for the post-selected outcomes of $\hat{M}_{\uparrow\uparrow}$, $\hat{M}_{\uparrow\downarrow}$, $\hat{M}_{\downarrow\uparrow}$ and $\hat{M}_{\downarrow\downarrow}$ are provided by
\begin{eqnarray}
 \ket{\psi}_{2,w}^{\hat{M}_{\uparrow\uparrow}} &=& \frac{1}{\sqrt{2}}(\ket{2,2}-\ket{0,0}),\nonumber\\
 \ket{\psi}_{2,w}^{\hat{M}_{\uparrow\downarrow}} &=& \frac{1}{\sqrt{2}}(\ket{2,0}+\ket{0,-2}),\nonumber\\
 \ket{\psi}_{2,w}^{\hat{M}_{\downarrow\uparrow}} &=& \frac{1}{\sqrt{2}}(\ket{0,2}+\ket{-2,0})\;,\nonumber\\
 \ket{\psi}_{2,w}^{\hat{M}_{\downarrow\downarrow}} &=& \frac{1}{\sqrt{2}}(\ket{0,0}-\ket{-2,-2}).
 \label{eq20}
\end{eqnarray}
It can be seen that if we post-select with the symmetric measurement operators ($\hat{M}_{\uparrow\uparrow}$ or $\hat{M}_{\downarrow\downarrow}$), the state of the two walkers is symmetric under the exchange. In fact, according to Theorem \ref{t1}, the state of the walkers after any $n^{th}$ step is symmetric as long as one post-selects the outcomes corresponding to these two operators. In general, a symmetric bipartite state is of the form
\begin{eqnarray}
    \ket{\xi}=\sum_{i \neq j}\big[\frac{\alpha^{\prime}_{ij}}{2}(\ket{a_i,a_j}+\ket{a_j,a_i})+\beta^{\prime}_{i}\ket{a_i,a_i}\big],\nonumber \\
    \label{25}
\end{eqnarray}
where $\{\ket{a_i}\}$ form an orthonormal basis and $\alpha^{\prime}_{ij}=\alpha^{\prime}_{ji}$. However, with Theorem \ref{t2}, the reduced density matrix of one walker has at most rank 2. This means at most only one $\alpha^{\prime}_{ij}$ and two $\beta^{\prime}_i$ are non-zero. Without loss of generality, Eq.~\ref{25} may be written as
\begin{eqnarray}
\ket{\xi}=\frac{\alpha^{\prime}_{ab}}{2}(\ket{a,b}+\ket{b,a}) + \beta^{\prime}_a\ket{a,a}+ \beta^{\prime}_b\ket{b,b}.\nonumber \\
\end{eqnarray}
In general, the state of this form would not be a Bell state except for the following two special cases,
\begin{itemize}
    \item If $\beta^{\prime}_{a}=\beta^{\prime}_{b}=0$ \\The two walkers completely occupy different space. The state becomes $\frac{1}{\sqrt 2}(\ket{a,b}+\ket{b,a})$ which is equivalent to the Bell state $\ket{\Psi^+}$.
    \item If $\alpha^{\prime}_{ab}=0$ and $|{\beta^{\prime}_a}|=|{\beta^{\prime}_b}|$\\ The state becomes $\frac{1}{\sqrt 2}(\ket{a,a}\pm\ket{b,b})$ which is equivalent to Bell state $\ket{\Phi^\pm}$.
\end{itemize}
Theoretically, since there is no other constraint that will guarantee either of the above condition is always satisfied, it is very unlikely in general that the entanglement between the walkers will always be constant at maximum. This is confirmed numerically as shown in Fig.~\ref{fig2}(a) for ($\hat{M}_{\uparrow\uparrow}$,$\hat{M}_{\downarrow\downarrow}$) post-selection outcomes.

Let us now consider the other two possible post-selections that are corresponding to the measurement operators, ($\hat{M}_{\uparrow\downarrow}$,$\hat{M}_{\downarrow\uparrow}$). Similar to the \textit{ACS} cases, the state of the two walkers cannot be a Bell state and the entanglement would not be constant as shown in Fig.~\ref{fig2}(b). Our numerical results showed that the entanglement between the two walkers for $\hat{M}_{\uparrow\uparrow}$ and $\hat{M}_{\uparrow\downarrow}$ oscillates and approaches asymptotic values of $0.7563$ and $0.7454$, respectively.

Our analysis for the \textit{SCS} case, so far, shows that, unlike \textit{ACS}, no post-selection with respect to the measurement basis $\{\hat{M}_{\uparrow\uparrow},\hat{M}_{\uparrow\downarrow},\hat{M}_{\downarrow\uparrow},\hat{M}_{\downarrow\downarrow}\}$ yields constant entanglement. In fact, this is also true for any local measurement basis because the only way one can obtain constant spatial entanglement is when the reduced state of the walkers become a Bell state whenever they are measured and post-selected. However, that is only guaranteed if the final walker states are always anti-symmetric ($\ket{\Psi^-}$). Given that the initial state of the whole system is symmetric, this is achievable if and only if one can post-select an anti-symmetric coin state. Since no local measurement could collapse the state being measured into an anti-symmetric state, it is impossible for the reduced state of the walkers to always be equivalent to $\ket{\Psi^-}$. In a sense, anti-symmetry really is a non-local property.

To complete this discussion, let us change from the local basis to the triplet-singlet basis. This, again, replaces two local measurement operators ($\hat{M}_{\uparrow\downarrow},\hat{M}_{\downarrow\uparrow}$) in the previous case with two non-local ones ($\hat{M}_{+}, \hat{M}_{-}$), hence it is now possible to post-select an anti-symmetric result. Note again that the only reason we allow such a non-local process for now is just because we would like to study how the final coin symmetry affect the entanglement. %The state of the two walkers is always symmetric for $\hat{M}_{+}$ post-selection and anti-symmetric for $\hat{M}_{-}$ post-selection. 
Next, let us consider, for example, the post-selection outcomes of $\hat{M}_{+}$ and $\hat{M}_{-}$ after the second step of the quantum walk given respectively as
\begin{eqnarray}
 \ket{\psi}_{2,w}^{\hat{M}_{+}} &=& \frac{1}{2}\big[\{(\ket{2}+\ket{-2})\otimes\ket{0}\}+\{\ket{0}\otimes(\ket{2} \nonumber\\
 &&+\ket{-2})\}\big]\;, \nonumber\\
 \ket{\psi}_{2,w}^{\hat{M}_{-}} &=& \frac{1}{2}\big[\{(\ket{2}-\ket{-2})\otimes\ket{0}\}-\{\ket{0}\otimes(\ket{2} \nonumber\\
 &&-\ket{-2})\}\big].\nonumber\\
 \label{32}
\end{eqnarray}
We can see that the walkers state is symmetric for $\hat{M}_{+}$ and the entanglement dynamics is oscillating and approach an asymptotic value of $0.5813$ as shown numerically in Fig.~\ref{fig2}(c). On the other hand, post-selection with the singlet basis $\hat{M}_{-}$ forces the state of the walkers to be anti-symmetric. The state of the walkers become $\ket{\Psi^-}$ now, similar to the \textit{ACS} with post-selected in triplet outcomes and the entanglement between the two walkers is constant as shown in Fig.~\ref{fig2}(d).

Let us now consider the other two initial states which are also \textit{SCS}, 
\begin{eqnarray}
\ket{\psi}_0=\ket{\Phi^{\pm}}\otimes\ket{0,0}.
\end{eqnarray}
After we perform quantum walk for $n^{th}$ steps, the state $\ket{\psi}_{n}$ is also symmetric. Similar to the $\ket{\Psi^+}$ case above, if we post-select with the operators ($\hat{M}_{\uparrow\uparrow}$,$\hat{M}_{\downarrow\downarrow}$,$\hat{M}_{+}$), the state of the two walkers becomes symmetric under exchange in accordance with Theorem \ref{t1}. We observe oscillating behaviour in the entanglement dynamics between the walkers for these post-selections. However, if we post-select with operator $\hat{M}_{-}$, the state of the two walkers become anti-symmetric (i.e. $\ket{\Psi^-}$) and the entanglement is not damping.

From the analysis of the \textit{SCS} cases above, we finally reach a conclusion that the condition for an undamping spatial entanglement is really when the final states of the walkers are anti-symmetric. Equivalently, this means the undamping behaviour will only occur if the symmetries between the initial and final coin states are opposite.
%%%%%%%%%%%%%%%%%%%%%%%%%%%%%%%%%%%%%%%%%%%%%%%%%%%%%%%%%%%%%%%%%%%%%%%%%%%%%%%%%
\subsubsection{Oscillation period of the spatial entanglement}
\begin{figure}[ht]
\center
\includegraphics[width=1\columnwidth]{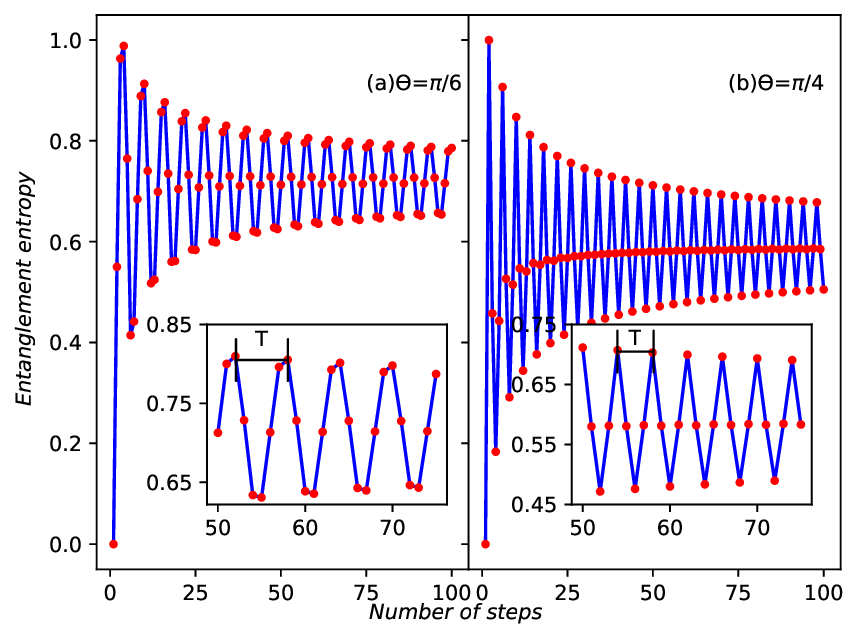}
\caption{\label{fig4} Entanglement between two walkers: (a) \textit{SCS} with $\hat{U}(\pi/6)$ for $\hat{M}_{\uparrow\uparrow}$ post-selection outcome: each repeated pattern contains 6 points and 1 cycle, and (b) \textit{ACS} with $\hat{U}(\pi/4)$ for $\hat{M}_{\uparrow\downarrow}$ post-selection outcome: each repeated pattern contains 4 points and 1 cycle.}
\end{figure}

By observing the numerical results in all the oscillating cases, both for the \textit{SCS} and \textit{ACS}, we discover that there is a relationship between the period per the repeated pattern of the entanglement dynamics $(T)$ and the angle parameter of the coin operator $(\theta)$ which is given by
\begin{eqnarray}
T=n\pi/\theta
\end{eqnarray}
where $n$ is the number of cycles per repeated pattern as shown Fig.~\ref{fig4}(a) for \textit{SCS} and Fig.~\ref{fig4}(b) for \textit{ACS}. A theoretical explanation for this behaviour remains an open question, but future work might reveal the underlying mechanism.
%%%%%%%%%%%%%%%%%%%%%%%%%%%%%%%%%%%%%%%%%%%%%%%%%%%%%%%%%%%%%%%%%%%%%%%%%%%%%%%%%
\section{\label{conc}Conclusion}
We present our study of the behaviour of the spatial entanglement between two walkers, given the following conditions: 
%the behaviour of spatial entanglement between two walkers, given the following conditions: 
\begin{enumerate}
    \item The initial states of the two walkers are pure and identical.  \label{condition 1}
    \item Each walker undergoes an identical discrete-time quantum walk process with respect to its corresponding coin.  \label{condition 2}
    \item The combined state of the two coins is initially bosonic (symmetric coin state: \textit{SCS}) or fermionic (anti-symmetric coin state: \textit{ACS}).  \label{condition 3}
    \item Only local measurements are allowed.  \label{condition 4}
\end{enumerate}
Firstly, we find that the state of the individual walker after its corresponding coin is measured and post-selected is always equivalent to a qubit at any step after the walk starts, i.e. effectively being two-dimensional. Secondly, we show that, if the symmetries between the initial and the final states of the coins are opposite, then the combined state of the two walkers is always equivalent to the singlet Bell state $\ket{\psi^-}$ and the initial coin entanglement is completely transferred to the spatial degrees of freedom. However, since only local measurements are allowed, only the \textit{ACS} case would yield such a result because it is physically impossible to implement an anti-symmetric local measurement as required for the \textit{SCS} cases. For the cases where the initial and the final symmetries between the two coins are the same and the cases where there is no symmetry between the final coin states, the spatial entanglement will fluctuate in an underdamping manner, gradually reaching an asymptotic value at a long time limit. We also numerically discover the relationship between the period per repeated pattern of the spatial entanglement ($T$) and the angle parameter of the coin operators ($\theta$) as follows, $T = n\pi/\theta$, where $n$ is the number of cycles per pattern.

In fact, the validity of our results extend beyond the discrete quantum random walk scenarios we considered in the main text, as long as the system of interest is equivalent to two spin-1/2 particles, satisfies Conditions $1$, $3$ and $4$ above, and each subsystem undergoes an independent and identical local unitary process relative to its counterpart. Therefore, our results, especially on the complete transfer and unchanging of the entanglement, may be useful for quantum communication and quantum cryptography applications because constant entanglement ensures a reliable and secure quantum information transfer over long distances.
\section{Acknowledgements}
The first author was supported by the ``Petchra Pra Jom Klao Ph.D. Research Scholarship from King Mongkut’s University of Technology Thonburi" Contract No.\,14/2561. This research has received funding support from the NSRF via the Program Management Unit for Human Resources \& Institutional Development, Research and Innovation [grant number B37F660011] and is supported by Thailand Science Research and Innovation (TSRI). Basic Research Fund: Fiscal year 2023 under project number FRB660073/0164. The researchers would like to thank Dr. Ekkarat Pongophas for inspiring discussions and useful suggestions. The researchers also acknowledge Innosoft, KMUTT for allowing access to their high-performance computing resources.
%%%%%%%%%%%%%%%%%%%%%%%%%%%%%%%%%%%%%%%%%%%%%%%%%%%%%%%%%%%%%%%%%%%%%%%%%%%%%%%%%%%%%%%%
\bibliographystyle{unsrtnat}
\bibliography{references}
\end{document}